\documentclass[english,aps,prb,twocolumn,superscriptaddress,floatfix,morefloats]{revtex4}
\usepackage[T1]{fontenc}
\usepackage[latin9]{inputenc}
\usepackage{verbatim}
\usepackage{amsmath}
\usepackage{amssymb}
\usepackage{graphicx}
\usepackage{times}

\makeatletter

\providecommand{\tabularnewline}{\\}

\@ifundefined{textcolor}{}
{%
 \definecolor{BLACK}{gray}{0}
 \definecolor{WHITE}{gray}{1}
 \definecolor{RED}{rgb}{1,0,0}
 \definecolor{GREEN}{rgb}{0,1,0}
 \definecolor{BLUE}{rgb}{0,0,1}
 \definecolor{CYAN}{cmyk}{1,0,0,0}
 \definecolor{MAGENTA}{cmyk}{0,1,0,0}
 \definecolor{YELLOW}{cmyk}{0,0,1,0}
}

\usepackage{tikz}
\usepackage{ifthen}
\usepackage{H1}

\usetikzlibrary{calc}
\usetikzlibrary{shapes.geometric}
\tikzset{hexagon/.style={regular polygon, regular polygon sides = 6}}

\newif\ifHexgridTriangulate
\newif\ifHexgridStartShifted
\pgfqkeys{/hexgrid}
         { name/.store in       = \HexgridName
         , xpos/.store in       = \HexgridX
         , ypos/.store in       = \HexgridY
         , rows/.store in       = \HexgridRows
         , cols/.store in       = \HexgridCols
         , size/.code           = {\pgfmathsetmacro{\HexDiameter}{#1}}
         , triangulate/.is if   = HexgridTriangulate
         , start shifted/.is if = HexgridStartShifted }

\tikzset{ every hexgrid hex/.style 2 args   = {draw}
        , every hexgrid triangulator/.style = {}}

\newcommand{\hexgrid}[2][]{
  \pgfqkeys{/hexgrid}{ name  = hexgrid , size = 1cm
                     , xpos  = 0       , ypos = 0
                     , triangulate   = false
                     , start shifted = false
                     ,#2 }

  \ifHexgridStartShifted
    \def\HexShiftModCheck{0}
  \else
    \def\HexShiftModCheck{1}
  \fi

  \begin{scope}[xshift=\HexgridX, yshift=\HexgridY,#1]
    \pgfmathsetmacro{\HexRadius}{\HexDiameter/2}
    \pgfmathsetmacro{\HexSide}{sqrt(3)*\HexRadius/2}
    \pgfmathsetmacro{\HexWidth}{2*\HexSide}

    \tikzset{every node/.style={hexagon, minimum size=\HexDiameter}}

    \foreach \row in {1,...,\HexgridRows} {
      \foreach \col in {1,...,\HexgridCols} {
        \pgfmathsetmacro{\HexX}%
                        {\HexWidth*(  (\col-1)
                                    + (mod(\row,2) == \HexShiftModCheck
                                        ? 0 : .5))}
        \pgfmathsetmacro{\HexY}%
                        {-(\HexRadius + \HexSide/2 + 2*\pgflinewidth)*(\row-1)}
        \node [hexagon, rotate=90, every hexgrid hex = {\row}{\col}]
              (\HexgridName-\row-\col)
              at (\HexX pt ,\HexY pt)
              {} ;
      }
    }

    \ifHexgridTriangulate
      \begin{scope}[every path/.style={every hexgrid triangulator}]
        \foreach \row in {1,...,\HexgridRows} {
          \foreach \col in {1,...,\HexgridCols} {
            \pgfmathparse{int(\row-1)}\let\prow\pgfmathresult
            \pgfmathparse{int(\col-1)}\let\pcol\pgfmathresult

            \ifnum\prow>0
              \draw    (\HexgridName-\prow-\col.center)
                    -- (\HexgridName-\row-\col.center) ;
            \fi
            \ifnum\pcol>0
              \draw    (\HexgridName-\row-\pcol.center)
                    -- (\HexgridName-\row-\col.center) ;
            \fi
            \ifnum\prow>0\ifnum\pcol>0
              \pgfmathparse{mod(\prow,2) == \HexShiftModCheck}
              \ifnum\pgfmathresult=1
                \draw    (\HexgridName-\prow-\col.center)
                      -- (\HexgridName-\row-\pcol.center) ;
              \else
                \draw    (\HexgridName-\prow-\pcol.center)
                      -- (\HexgridName-\row-\col.center) ;
              \fi
            \fi\fi
          }
        }
      \end{scope}
    \fi
  \end{scope}
}

\makeatother

\usepackage{babel}
\begin{document}

\preprint{This line only printed with preprint option}

\title{Numerical study of a transition between $\mathbb{Z}_{2}$
topologically ordered phases}

\author{Siddhardh C. Morampudi}

\affiliation{Max-Planck-Institut f\"ur Physik komplexer Systeme, Dresden, Germany}

\author{Curt von Keyserlingk}

\affiliation{Rudolf Peierls Centre for Theoretical Physics, 1 Keble Road, Oxford,
OX1 3NP, United Kingdom}

\author{Frank Pollmann}

\affiliation{Max-Planck-Institut f\"ur Physik komplexer Systeme, Dresden, Germany}
\begin{abstract}
Distinguishing different topologically ordered phases and characterizing
phase transitions between them is a difficult task due to the absence
of local order parameters. In this paper, we use a combination of
analytical and numerical approaches to distinguish two such phases
and characterize a phase transition between them. The ``toric code''
and ``double semion'' models are simple lattice models exhibiting $\mathbb{Z}_{2}$
topological order. Although both models express the same topological
ground state degeneracies and entanglement entropies, they are distinct
phases of matter because their emergent quasi-particles obey different
statistics. For a 1D model, we tune a phase transition between these
two phases and obtain an exact solution to the entire phase diagram,
finding a second-order Ising$\times$Ising transition. We then use
exact diagonalization to study the 2D case and find indications of
a first-order transition. We show that the quasi-particle statistics
provides a robust indicator of the distinct topological orders throughout
the whole phase diagram.
\end{abstract}
\maketitle

\section{introduction}

Phases of matter and phase transitions are usually classified using
Landau's symmetry breaking theory, which distinguishes different phases
of matter using local order parameters. However, it has become clear that 
some quantum phases do not fall into this paradigm; they posses a 
\emph{topological order}\cite{WEN1990} which cannot be detected by local order parameters.
Prominent examples of these phases include the fractional quantum
Hall effect\cite{Laughlin1983} and gapped spin liquids\cite{Wen1991,Sachdev1991,Moessner2001,Balents2010,Yan2011,Han2012}
that appear in some highly frustrated systems. 

Topological order is characterized by presence of \emph{long-range entanglement},
which means that the state cannot be transformed into a product state through a 
local unitary evolution.\cite{Chen2010} This long-range
entanglement can give rise to some unique properties in topologically
ordered phases such a topological ground state degeneracy and fractionalized
(anyonic) excitations.\cite{Nayak2008} The topological degeneracy
is a degeneracy of the ground-state wavefunction that depends on the
topology of the system and is robust against any small and local perturbations.
The anyonic excitations are characterized by their braiding statistics
that is quantified by two matrices, the $U$ and $S$-matrices. The two
matrices correspond to the exchange and mutual statistics of the excitations
respectively.\cite{KESKI-VAKKURI1993} 

Since topological order cannot be characterized by local order parameters,
it is difficult to uniquely identify a topologically ordered phase. The most
common method to identify a topologically ordered phase is through
the use of the topological entanglement entropy.\cite{Kitaev2006,Levin2006,Furukawa2007,Isakov2011,Jiang2012,Grover2013}
However, this method is not unique as multiple topologically ordered phases
can have the same topological entanglement entropy.\cite{Wen} A more
unique characterization is to directly calculate the $U$ and $S$-matrices of the
phase. This approach has been used to detect topologically ordered phases in a 
number of recent works.\cite{Zhang2012,Cincio2013,Zaletel2013,Zhu2013} Another interesting question is the nature of phase transitions
between topologically ordered phases. Since these transitions do not
fall readily into the traditional symmetry-breaking paradigm, they are 
still poorly understood in general. Previous numerical studies 
have so far mostly concentrated on transitions from topologically ordered phases into trivial phases
and not between two topologically ordered phases.\cite{Trebst2007,Castelnovo2008}

To explore the above questions, we consider two microscopic Hamiltonians,
the toric code (TC)\cite{Kitaev2003} and the double semion (DSem)\cite{Freedman2004,PhysRevB.71.045110}
model. These are simple examples of more general string-net models
which describe a large class of achiral topologically ordered phases.
The TC model is a canonical example of topological order in 2D; it
possesses a ground state degeneracy on a torus and non-vanishing topological
entanglement entropy which distinguishes it from a trivial product state.
The TC order is found in superconductors\cite{Hansson2004} and the
Heisenberg antiferromagnet on the kagome lattice.\cite{Yan2011,Jiang2012,Depenbrock2012}
The related DSem model exhibits
the same degeneracy and entanglement entropy as the TC model, but
represents a different kind of order because its excitations have
different statistics. The DSem phase can be realized as a bilayer of $\nu=\pm1/2$ 
bosonic Laughlin-states,\cite{Grover2013}  and may also be realized in spin-models on the kagome lattice.\cite{Barkeshli2013,Bauer2014}
Both the TC and the DSem represent $\mathbb{Z}_{2}$ topological orders 
because they are both $\mathbb{Z}_{2}$ gauge theories.\cite{Dijkgraaf1990,Bais1993}

The aim of this paper is two-fold. First, we examine a Hamiltonian
that exhibits multiple topologically ordered phases and we
try to identify its phase diagram. We make use of recent advances
that allow us to calculate the $U$ and $S$-matrices numerically from a
full set of topologically degenerate ground state wave functions.\cite{Zhang2012,Cincio2013,Zhu2013}
We implement the procedure using exact diagonalization and show that,
even for small systems, the $U$ and the $S$-matrices give a far more robust
diagnostic of the topological order of a phase than the more traditional
topological entanglement entropy measurement. Second, we discuss the
nature of the transition between the TC and DSem phases. In phase
transitions between the TC and non-topological states, the disappearance
of topological order is due a to a condensation of vortices, which
leads to a confinement of point charges.\cite{Trebst2007,Castelnovo2008,Schmidt2011}
However, the tuning between TC and DSem models cannot
be viewed as a simple confinement transition. Using exact diagonalization,
we find indications for a first-order transition at the boundary of
the two phases. 

This work is organized as follows. In Sec. \ref{sec:model}, we review
the Hamiltonians for the TC and DSem models, and give a natural prescription
for tuning between the two models. This gives rise to a family of
Hamiltonians which we solve exactly for a 1D chain in Sec. \ref{sec:1D}.
In Sec. \ref{sec:two-dimensions}, we perform an exact diagonalization
study of the transition between the TC and DSem models for a 2D system.
We first extract the nature of the transition using the energy spectrum
and then calculate the $U$ and the $S$-matrices numerically which we use
as non-local order parameters to distinguish the two phases. We
finally conclude in Sec. \ref{sec:Summary-and-discussions} with a
summary and brief discussion of our main results.

\section{model\label{sec:model}}

\begin{figure}
\includegraphics[width=0.99\columnwidth]{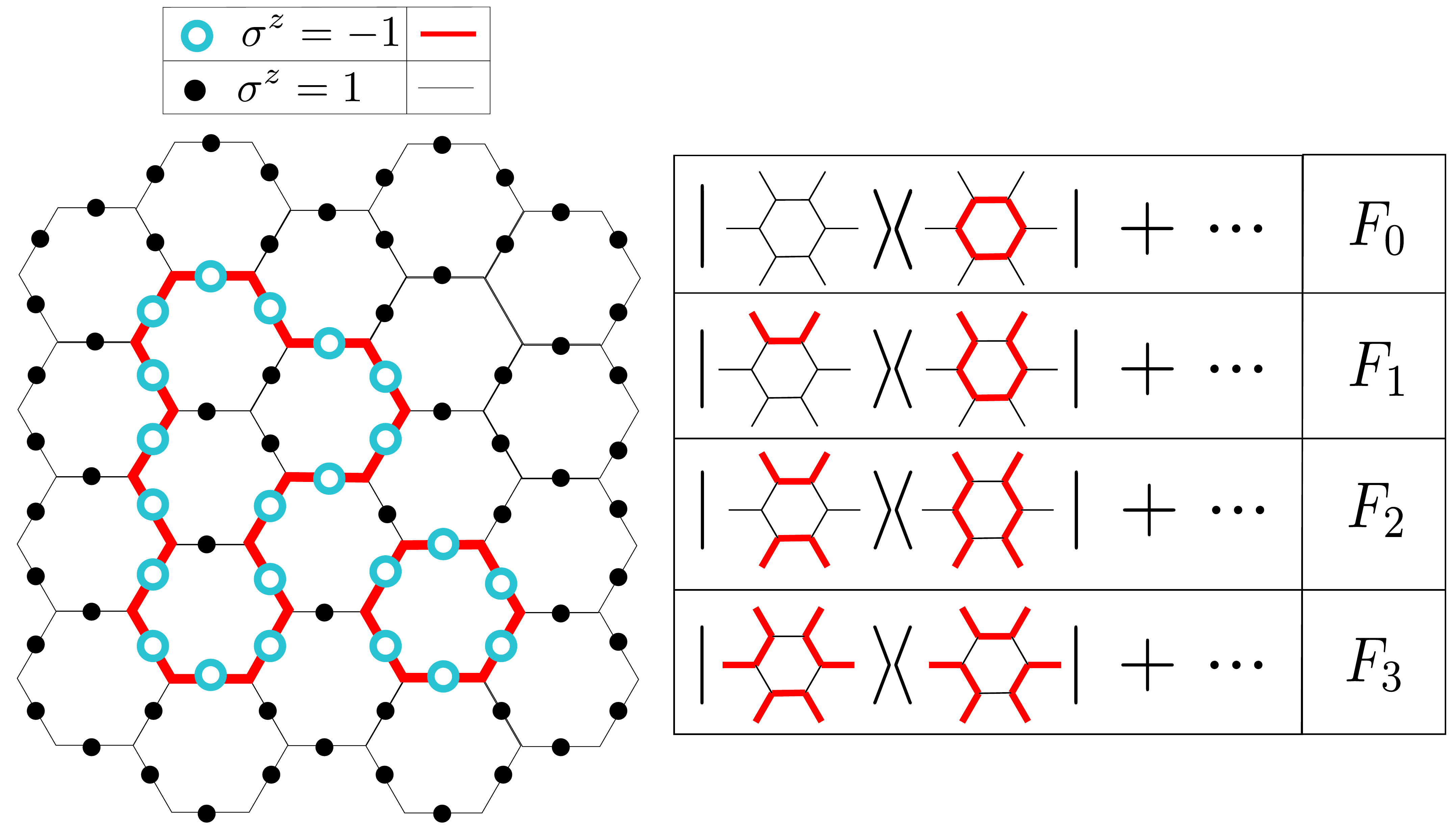}

\begin{tabular}{cc}
(a) & ~~~~~~~~~~~~~~~~~~~~~~~~~~~~~~~~~~(b)\tabularnewline
\end{tabular}\caption{(a) indicates the mapping from the spin to the loop space. (b) shows
the terms in the Hamiltonian. The dots indicate hermitian conjugates
of the ones shown here and also equivalent terms obtained by rotating the
lattice.}
\label{LoopsHam}

\end{figure}

Before introducing our model, we give a brief review of the TC and
DSem models,\cite{Freedman2004,PhysRevB.71.045110} their ground states,
and the statistics of their emergent excitations.

\subsection*{Toric Code (TC)}

The Hilbert space of the TC consists of a tensor product of $\sigma^{z}=\pm1$
living on each edge of a honeycomb lattice. In the remainder of this
paper, we will visualize configurations of these spins by coloring
in edges with $\sigma^{z}=-1$, while leaving $\sigma^{z}=1$ uncolored
 as indicated in Figure \ref{LoopsHam}(a). The Hamiltonian has the form

\begin{equation}
H^{\mathrm{TC}}=-\epsilon_{P}\sum_{p}\underbrace{\prod_{k\in\partial p}\sigma_{k}^{x}}_{B_{p}^{\mathrm{TC}}}-\epsilon_{V}\sum_{v}\underbrace{\prod_{k\in s(v)}\sigma_{k}^{z}}_{B_{v}}
\end{equation}
with $\epsilon_{P},\epsilon_{V}>0$. The sums $\sum_{p}$, $\sum_{v}$ run over all plaquettes and
vertices respectively, $\partial p$ are the six edges on the boundary
of plaquette $p$, and $s(v)$ are the three edges attached to vertex
$v$. The vertex operator $B_{v}$ has eigenvalues $\pm1$ depending on whether
there are an even/odd number of down-spins on the edges coming into
vertex $v$. On the other hand, the plaquette term $B_{p}^{\mathrm{TC}}$ flips
the spins on every edge of a plaquette $p$ and commutes with $B_{v}$
because it flips a pair of spins at each vertex $v$. Since also clearly
$\left[B_{v},B_{v'}\right]=\left[B_{p}^{\mathrm{TC}}B_{p'}^{\mathrm{TC}}\right]=0$, the Hamiltonian
is a sum of commuting operators, allowing us to write down its ground
states and excitations exactly. The ground states are defined by $B_{p}^{\mathrm{TC}}=B_{v}=1$
for all $v,p$. The $B_{v}=1$
condition implies that the ground state is a superposition of closed
loop configurations of the $\sigma^{z}$ spins, while the $B_{p}^{\mathrm{TC}}=1$
condition implies that all terms in this superposition have equal
weight. 

When the system is placed on a torus, the $\mathbb{Z}_2$ topological order of the TC model manifests itself through a four-fold degeneracy.
The degenerate ground states can be distinguished by the winding number parities around the torus measured by
\begin{equation}
P^z_{\gamma_{x(y)}}=\prod_{k\in \gamma_{x(y)}}\sigma_{k}^{z} , 
\end{equation}
where $\gamma_{x(y)}$ is a long loop around the $x (y)$ direction of the torus (Fig.~\ref{TorusPartition}). We then label the four sectors as (00, 01, 10 and 11), according to whether there is an even/odd number of loops winding around the $x$ and $y$ direction of the torus.

For future reference we note that we can rewrite the plaquette term (at least when all $B_{v}=1$, i.e., no broken loops are allowed)
as 
\begin{equation}
B_{p}^{\mathrm{TC}}=F_{0}+F_{1}+F_{2}+F_{3}
\end{equation}
\noindent where the $F$'s are defined in Figure \ref{LoopsHam}(b).
The operator $F_{0}$ either creates or destroys a single loop, $F_{1}$ changes
the size of the loops, $F_{2}$ changes the number of loops by joining
or breaking them and $F_{3}$ changes the number of loops by an even
number.

There are three non-trivial deconfined anyon charges in the theory:
the charge $1$ electric defects $e$, the $\pi$ flux magnetic defects
$m$, and a bound-state of charge and flux $\psi\equiv e\times m$
which behaves like a fermion. In the model, an $e$ particle is associated
with a vertex defect $B_{v}=-1$, the $m$ particle is associated
with a plaquette defect $B_{p}^{\mathrm{TC}}=-1$ and the fermion $\psi$ is
associated with both a vertex and plaquette defect. The excitations
can only be created in pairs. For example, breaking the end of a loop
creates two electric defects ($e$) at the two ends of the broken
loop.

\subsection*{Double Semion (DSem)}

The DSem model is defined on the same Hilbert space as the TC, and
has the Hamiltonian 

\begin{equation}
H^{\mathrm{DSem}}=\epsilon_{P}\!\!\sum_{p}\underbrace{\prod_{k\in\partial p}\sigma_{k}^{x}\!\!\left[\prod_{j\in s(p)}i^{(1-\sigma_{j}^{z})/2}\right]}_{B_{p}^{\mathrm{DSem}}}\proj_{v}-\epsilon_{V}\!\!\sum_{v}\underbrace{\prod_{k\in s(v)}\sigma_{k}^{z}}_{B_{v}}
\end{equation}

\noindent where $s(v),\partial p$ are defined as before, and where
$s(p)$ is the set of six legs radiating from plaquette $p$, and
$\proj_{v}$ is a projector onto states with $B_{v}=1$ for all $v$. The
DSem Hamiltonian is the same as that of the TC, except that the plaquette
terms differ: firstly, the plaquette term is only active on states
with $\proj_{v}=1$. Secondly, in addition to flipping the edges of
$p$, the new $B_{p}^{\mathrm{DSem}}$ operator has a phase of $\pm1$ depending on
the spins in $s(p)$. In addition, the plaquette component of the
DSem Hamiltonian appears with a total ($+$) sign rather than a ($-$)
sign.

It can be verified that $\left[B_{v},B_{p}^{\mathrm{DSem}}\right]=0$, so we
may restrict ourselves to the study of the subspace with all $B_{v}=1$,
i.e., we consider states consisting of superpositions of closed loops.
In this subspace, $\left[B_{p}^{\mathrm{DSem}},B_{p'}^{\mathrm{DSem}}\right]=0$ for all
plaquettes $p,p'$. Once again, the Hamiltonian is exactly solvable,
and the ground state saturates a lower bound on the Hamiltonian by
having $B_{p}^{\mathrm{DSem}}=-1$ and $B_{v}=1$ for all $p,v$.  Like the TC, the DSem 
model has a topological degeneracy when placed on a torus.

At this point, it is helpful to rephrase the plaquette term in the
$\proj_{v}=1$ sector slightly differently. We can rewrite the plaquette
term in the form:

\begin{equation}
B_{p}^{\mathrm{DSem}}=F_{0}-F_{1}+F_{2}-F_{3}
\end{equation}

\noindent Here we see that the $F_{0},F_{2}$ terms have a relative
sign with respect to the $F_{1},F_{3}$ terms -- in the TC case, all
the $F$ terms have a $+$ sign. $F_{0}$ and $F_{2}$ can change
the number of loops by an odd number and thus can change the parity
of the loop sector whereas $F_{1}$ and $F_{3}$ only change the number
of loops by an even number and thus preserve the parity of the loop
sector. This, along with the fact that $B_{p}^{\mathrm{DSem}}=1$ tells us that the
ground state is a superposition of loop configurations weighted with
a phase of $U=(-1)^{\#\mathrm{loops}}$. Indeed, this operator $\hat{U}$
is a non-local unitary transformation relating the two Hamiltonians
to each other (at least within the $\proj_{v}=1$ sector) through

\begin{equation}
H^{\mathrm{DSem}}=(-1)^{\#\mbox{loops}}H^{TC}(-1)^{\#\mbox{loops}}\label{eq:dualitytransform},
\end{equation}

\noindent and thus the two models have the same spectrum.
Like the TC, the DSem model has three non-trivial deconfined anyonic
charges. There are positive and negative chirality semions $s^{+},s^{-}$
as well as a $\pi$ flux defect which can be thought of as a bound
state of the semions $m=s^{+}\times s^{-}$. There is some freedom
and subtlety involved in defining these excitations near their endpoints,\cite{Burnell2013}
but for the purposes of this work the $s^{+}$ is associated with
a vertex defect $B_{v}=-1$, the $m$ is associated with a plaquette
defect $B_{p}^{\mathrm{DSem}}=1$ while the $s^{-}$ is associated with a combination
of plaquette and vertex defects. 

We have seen that a simple change in the definition of the plaquette
operators has led to a substantial change in the excitations
present in the theory. To explore a transition between these phases,
we consider a Hamiltonian which interpolates between the TC and DSem
model 

\begin{equation}
H(\eta,\mu,\lambda)=\eta H^{\mathrm{TC}}+\mu H^{\mathrm{DSem}}-\lambda\sum_{k=1}^{n}\sigma_{k}^{z}\label{eq:mainH}
\end{equation}

\noindent where we have also introduced a magnetic field term $\lambda$. 
We work in a basis with only closed loops (Setting $\epsilon_{V}$ 
to a large value to heavily penalize broken loops) and set $\epsilon_{P}=1$.
The magnetic field then acts as a string tension for the loops.
We also impose the condition $\eta+\mu+\lambda=1$ to explore a triangular 
phase diagram with exactly solvable points on the vertices. 
This Hamiltonian also has a topological degeneracy when placed on a torus.
For any $\eta$ or $\mu$, a sufficient increase in $\lambda$ will lead to a polarization
of spins $\sigma^{z}=1$. For $\eta=0$ and $\mu>0$ or $\eta>0$
and $\mu=0$, the transition to the polarized phase with increasing
$\lambda$ can be viewed as the standard confining transition in $(2+1)D$
$\mathbb{Z}_{2}$ gauge theory where $\pi$ flux vortices proliferate and destroy
the topological order. This gives rise to 3D Ising critical exponents.\cite{Trebst2007}
We will mainly be concerned with the transition associated with $\lambda=0$
and tuning $\mu$ from $0$ to $1$ (or $\eta$ from $1$ to $0$),
which \emph{is not a standard condensation transition }because both
sides of the transition have the same number of deconfined species\emph{. }

\section{The 1D transition\label{sec:1D}}

\begin{figure}
\begin{tabular}{cc}
a) & \tabularnewline
 & \includegraphics[width=0.95\columnwidth]{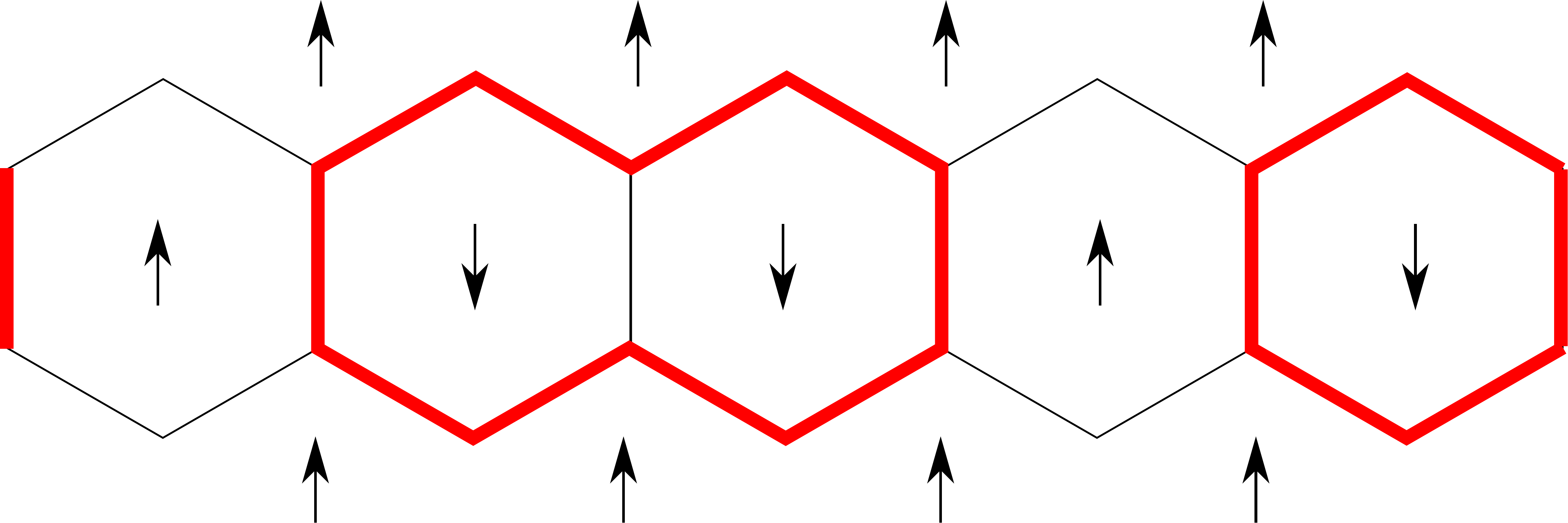}\tabularnewline
 & \tabularnewline
 & \tabularnewline
b) & \tabularnewline
 & \includegraphics[width=0.95\columnwidth]{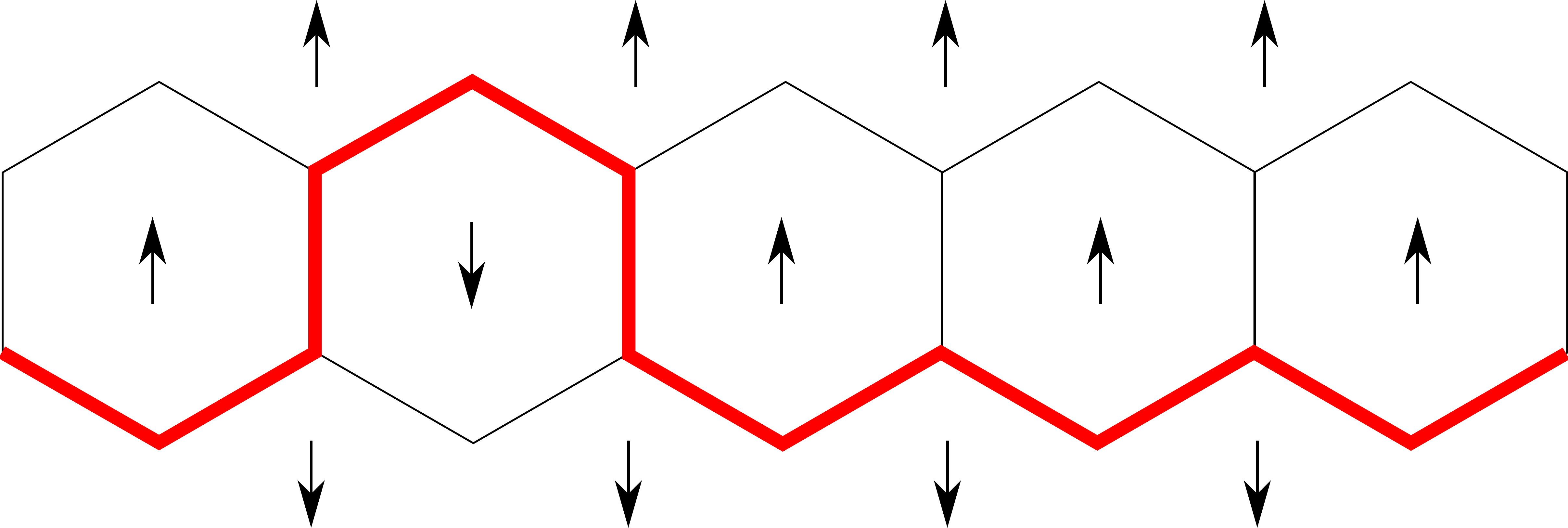}\tabularnewline
\end{tabular}

\caption{The ladder geometry used for the quasi-1D systems. In (a), we show
the domain wall representation of loops in the even sector of the
Hilbert space; note the use of auxiliary $\uparrow$ spins off the
lattice. (b) shows a domain wall representation for loops in the odd
sector of the Hilbert space; here we need auxiliary $\uparrow$ spins
above the chain and $\downarrow$ spins below it. }
\label{fig:ladder}
\end{figure}

\begin{figure}[!h]
\includegraphics[width=1\columnwidth]{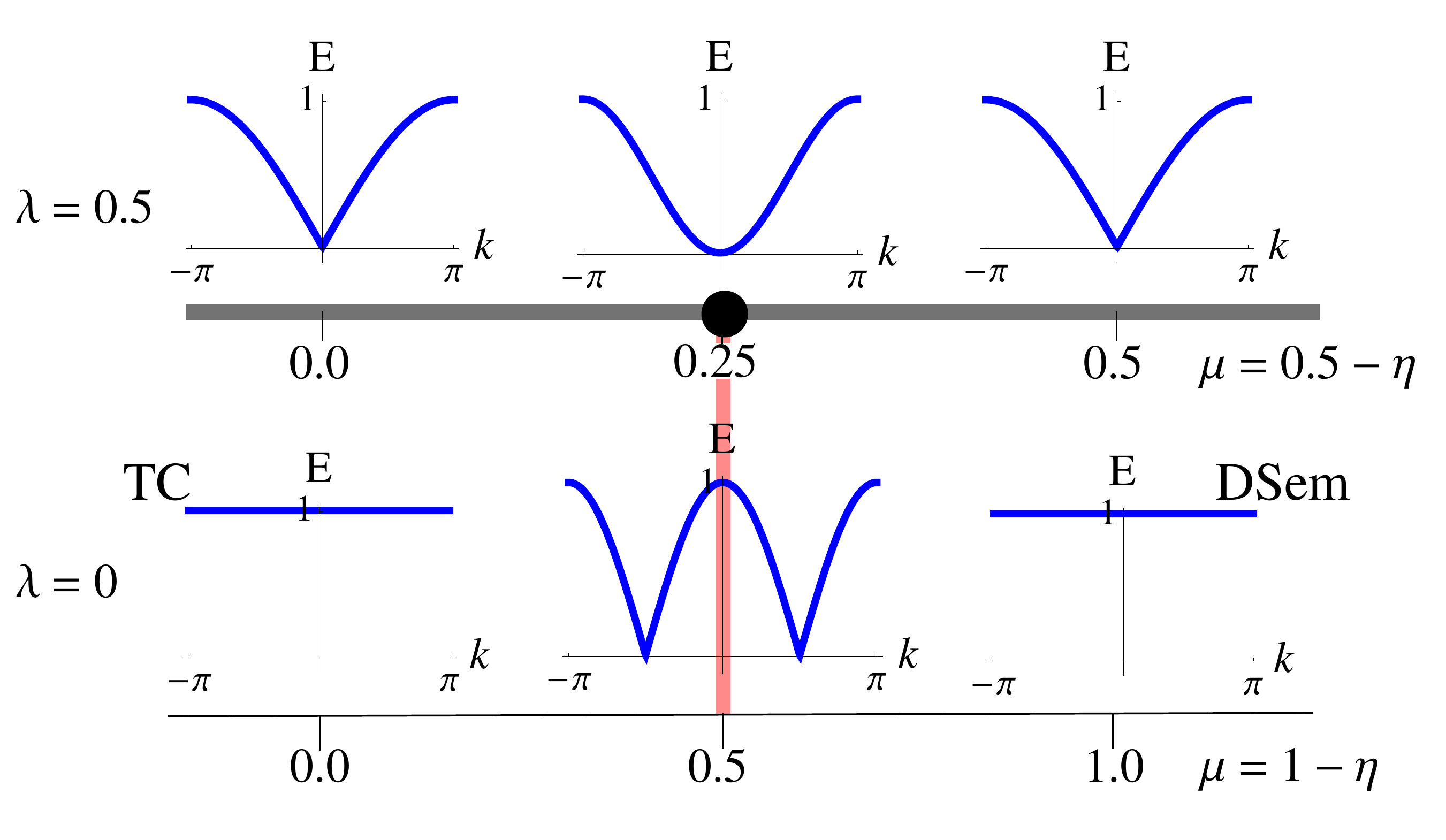}

\caption{The spectrum of the 1D ladder model in the even sector at various
points in the phase diagram. For $0\leq\eta<\mu$ ($0\leq\mu<\eta$)
and $0\leq\lambda<\mu(\eta)$ we are in the DSem (TC) phase, and the
spectrum is fully gapped. For any $0\leq\lambda<0.5$ there is a continuous
phase transition involving two critical Ising theories as we tune
$\mu$ to be greater than $\eta$ (vertical red line). For $\mu\neq\eta$ and when $\lambda=0.5$ (horizontal thick line), there is a critical point in the 2D critical Ising universality class, and the spectrum has a single Dirac point. For $\lambda>0.5$, $\sigma^z$ loops are short, and the auxiliary Ising spins $\tau^z$ are ordered. At $\lambda=0.5$,
$\mu=\eta=0.25$ which is the intersection of two lines of critical
points (horizontal and vertical), the spectrum shows a quadratic dispersion.}
\label{fig:ladderspectrum}
\end{figure}

Before discussing the 2D topological phase transition, we solve the
problem exactly on a quasi-1D ladder geometry\cite{Gils2009} (Fig.~\ref{fig:ladder})
with the Hamiltonian of Eq.~(\ref{eq:mainH}). We impose periodic boundary
conditions, and further restrict ourselves to the sector of the Hilbert
space with $B_{v}=1$ i.e., no vertex defects. Loop configurations
in this restricted space can be labelled even/odd, depending on whether
an even or odd number of loops wind around the system. We concentrate
on the even loop sector, and comment later on the odd loop sector
which is less interesting because it reduces to the TC ladder Hamiltonian.

In the even sector [Fig.~\ref{fig:ladder}(a)], loop configurations
can be realized as domains walls between Ising spins $\tau^{z}$ defined
at the center of plaquettes. We first introduce auxiliary spins into
the system: place $\tau^{z}=1$ spins above the ladder, and $\tau^{z}=1$
spins below the ladder as in Fig.~\ref{fig:ladder}. A string is present
on an edge ($M$) of the lattice precisely when the plaquettes $p,q$
bordering $M$ have different Ising spins, so that $\sigma_{M}^{z}=\tau_{p}^{z}\tau_{q}^{z}$.
Moreover, flipping the value of the Ising spin $\tau_{p}^{z}$ on
a plaquette corresponds to adding a loop around plaquette $p$, so
that $\tau_{p}^{x}=\prod_{M\in\partial p}\sigma_{M}^{x}$. With these
variables

\begin{equation}
H(\eta,\mu,\lambda)=-\sum_{r}\tau_{r}^{x}\left[\eta-\mu\tau_{r-1}^{z}\tau_{r+1}^{z}\right]-\lambda\sum_{r}\tau_{r}^{z}\tau_{r+1}^{z}\punc{.}\label{eq:1DHamiltonian}
\end{equation}

\noindent This effective Hamiltonian has been studied in the context of cold atoms.\cite{Pachos2004}
For $\eta=\mu$ and $\lambda=0$, this is also the effective
Hamiltonian for the edge of a 2D Ising Symmetry Protected Topological
(SPT) phase\cite{Levin2012} with Ising symmetric boundary conditions.
As an aside, we note that the odd winding number sector has the effective
Hamiltonian $H(\eta+\mu,0,\lambda)$ using the notation in Eq.~(\ref{eq:1DHamiltonian}).
This is just a TC Hamiltonian in a magnetic field. 

The general Hamiltonian Eq.~(\ref{eq:1DHamiltonian}) is amenable to
a Jordan-Wigner transformation using $\tau_{r}^{x}=1-2c_{r}^{\dagger}c_{r}$
and $\tau_{r}^{z}=\left\{ \prod_{s<r}(1-2c_{s}^{\dagger}c_{s})\right\} (c_{r}+c_{r}^{\dagger})$.
In the Majorana Fermion language $i\psi_{r}=c_{r}^{\dagger}-c_{r}$
and $\chi_{r}=c_{r}^{\dagger}+c_{r}$, and the Hamiltonian can be
rewritten as:

\begin{align}
H(\eta,\mu,\lambda)= & -i\sum_{{s}}\left[\eta\chi_{s}^{a}\psi_{s}^{a}+\mu\chi_{s+1}^{a}\psi_{s}^{a}\right]-i\lambda\sum_{r}\psi_{r-1}\chi_{r} \nonumber \\
 & -i\sum_{{s}}\left[\eta\chi_{s}^{b}\psi_{s}^{b}+\mu\chi_{s+1}^{b}\psi_{s}^{b}\right]
\end{align}

\noindent where $\chi_{s}^{a}=\chi_{2s}$ and $\chi_{s}^{b}=\chi_{2s+1}$
and similarly for $\psi$. Thus the system consists of two fermion
chains (the even and odd sites) coupled only by the magnetic field
term $\lambda$.

The spectrum is exactly symmetric about $\eta=\mu$, which follows
from the unitary transformation of Eq.~(\ref{eq:dualitytransform}).

Setting $\lambda=0$,  for $\eta= 0$ and $\mu =1$  ($\eta =1$  and $\mu = 0$), we at the DSem (TC) fixed point and the spectrum is perfectly flat as expected. In both cases, loops ($\sigma^{z}=-1$) proliferate,
or equivalently the $\tau^{z}$ Ising spins are disordered. Increasing
$\lambda$ away from either of the fixed points eventually leads to the confinement
of loops (corresponding to an ordering of the dual Ising spins). The
transition between short and long loop phases (confined and deconfined) is in 
the 2D Ising critical universality class. Setting $\lambda=0$ and $\eta=1-\mu$ and tuning
from $\mu=0$ to $1$ leads to a single transition at $\mu=\eta=1/2$
with two gapless points, corresponding to two decoupled Ising models (Fig.~\ref{fig:ladderspectrum}).
It is interesting to notice that, precisely at $\mu=\eta$, the number
of domain walls $\hat{Q}=\sum_{r}\frac{1+\sigma_{r}^{z}\sigma_{r+1}^{z}}{4}=\sum_{r}\frac{1+i\psi_{r}\chi_{r+1}}{4}$
is a good quantum number. This symmetry is apparent in the fermion
language if we rewrite 
\begin{align}
H(\eta,\mu,\lambda)= & -i\sum_{{r}}\tilde{\Psi}_{r+1}^{T}\left(\begin{array}{cc}
\mu & 0\\
0 & \eta
\end{array}\right)\Psi_{r}+\frac{i\lambda}{2}\sum_{r}\tilde{\Psi}_{r}^{T}\Psi_{r}
\end{align}

\noindent where $\Psi_{r}=\left(\begin{array}{cc}
\psi_{r-1} & \chi_{r}\end{array}\right)^{T}$, $\tilde{\Psi}_{r}=\left(\begin{array}{cc}
\chi_{r} & -\psi_{r-1}\end{array}\right)^{T}$. The U(1) symmetry emerges at $\eta=\mu$ and is associated with
the real $SO(2)$ rotation $\Psi\rightarrow\hat{R}\Psi$, and under
which $\tilde{\Psi}\rightarrow\hat{R}\tilde{\Psi}$. One striking
feature of the 2D problem (which we study next) at the transition
is that the U(1) conservation law of domain-walls/loops at $\eta=\mu$
is broken down to a $\mathbb{Z}_{2}$ conservation law of loop number.

\section{two dimensions\label{sec:two-dimensions}}

Having concluded the nature of the phase diagram and the phase transitions
in 1D, we examine the system in 2D on a $L\times L$ honeycomb lattice.
We use exact diagonalization to obtain the ground state and the lowest
excited states. The exponential growth of the Hilbert space with the
system size usually does not allow simulations beyond approximately
30-40 spins in exact diagonalization. However, restricting our Hilbert
space to the space of loops allows us to go up to 75 spins. 

We first obtain the energy spectrum from which we examine the nature
of transition between the TC and DSem phases. 
We then try to identify the topological and non-topological regions
in the phase diagram by looking at the behavior of the topological
entanglement entropy (TEE). We only calculate the spectrum and the TEE for
the topological sector with an even number of long loops around the torus (00 sector) 
as we expect the sectors be degenerate in the thermodynamic
limit. Having identified the topological regions, we characterize the phases
in the topological regions by extracting the braiding statistics of
their excitations.

\subsection{Energy Spectrum}

\begin{figure}
\centering{}\includegraphics[width=1\columnwidth]{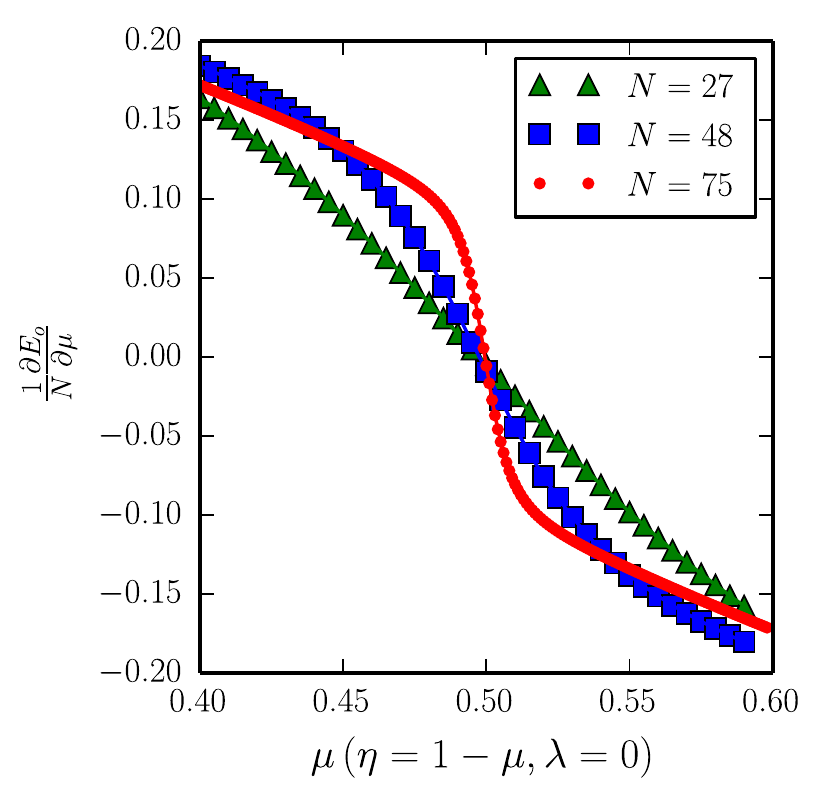}\caption{Derivative of ground state energy density as the system is tuned from the
TC to DSem phase. The increase in steepness at $\mu=\eta=0.5$
as we go to large system sizes ($N$ = number of spins) indicates
a possible first order transition.}
\label{GSEnergy}
\end{figure}

In Eq.~(\ref{eq:dualitytransform}) we noted that $H^{\mathrm{DSem}}$ and 
$H^{\mathrm{TC}}$ are related through a unitary transformation $U=(-1)^{\#\mathrm{loops}}$ 
and thus the spectrum is symmetric about $\mu=\eta$, regardless of $\lambda$ (because the 
string tension $\lambda$ does not change the number of loops). Figure \ref{GSEnergy} shows 
how the first derivative of the ground state energy density of the system changes as we
tune from the TC point to the DSem point at zero magnetic field. The derivative 
becomes steeper with increasing system size at $\mu = 0.5$. Extrapolating to the thermodynamic 
limit, we expect it to develop a discontinuity thus indicating a possible first-order transition

We can get additional information about the transition by studying
the behavior of the excitations near the transition point.
For example, at a second-order transition, a condensation of excitations
is expected. This motivates
the definition of a quantity which is related to the fidelity (the overlap
of the ground state wavefunctions at two neighboring points in the
transition). We consider  the spectral decomposition of the ground state at
the transition point $|\psi_{0}^{c}\rangle$ into the eigenstates of the Hamiltonian at neighboring points $x$.
From this we can directly see  which eigenstates contribute to the ground state $|\psi_{0}^{c}\rangle$ at the transition.
To be more precise, we define the following quantity,
\begin{equation}
A(\omega,x)=\sum_{n}|\langle\psi_{n}^{x}|\psi_{0}^{c}\rangle|\delta(\epsilon_{n}^{x}-\omega) \label{eq:sd}
\end{equation}

\noindent where $|\psi_{n}^{x}\rangle$ and $\epsilon_{n}^{x}$ are
the $n^{th}$ excited eigenstate and eigenvalue of the Hamiltonian
at $x$. It is difficult to obtain more than a few excited states $|\psi_n^x\rangle$ directly using efficient sparse matrix methods. 
We thus use instead an equivalent representation of (\ref{eq:sd}),

\begin{equation}
A(\omega,x)=\dfrac{1}{\pi}\textrm{Im}\langle\psi_{0}^{c}|\dfrac{1}{\omega-i\zeta-H^{x}}|\psi_{0}^{c}\rangle
\end{equation}

\noindent where $\zeta\rightarrow0$ and $H^{x}$ is the Hamiltonian
at $x$. This function can be evaluated efficiently by using the continued fraction method.\cite{Dagotto94} 

\begin{figure}
\noindent \begin{raggedright}
\begin{tabular}{cc}
~~~~~~~~~~~~~~~~~~~~~~~(a) & ~~~~~~~~~~~~~~~~~~~~~~~~~~~~~~~~~~~~~~~~~~(b)\tabularnewline
\end{tabular}
\par\end{raggedright}

\includegraphics[width=1\columnwidth]{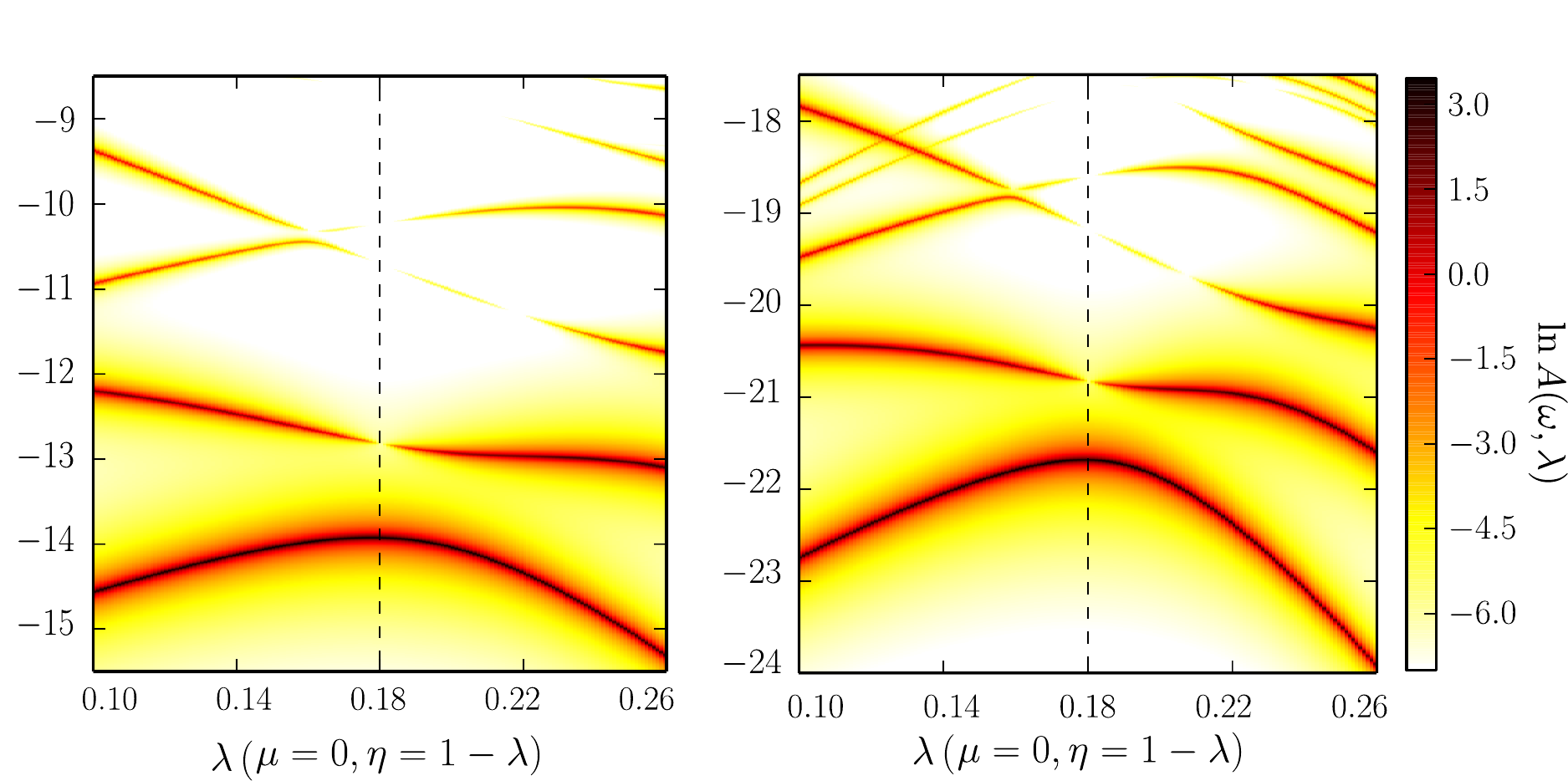}

\noindent \begin{raggedright}
\begin{tabular}{cc}
~~~~~~~~~~~~~~~~~~~~~~~(c) & ~~~~~~~~~~~~~~~~~~~~~~~~~~~~~~~~~~~~~~~~~~(d)\tabularnewline
\end{tabular}
\par\end{raggedright}

\includegraphics[width=1\columnwidth]{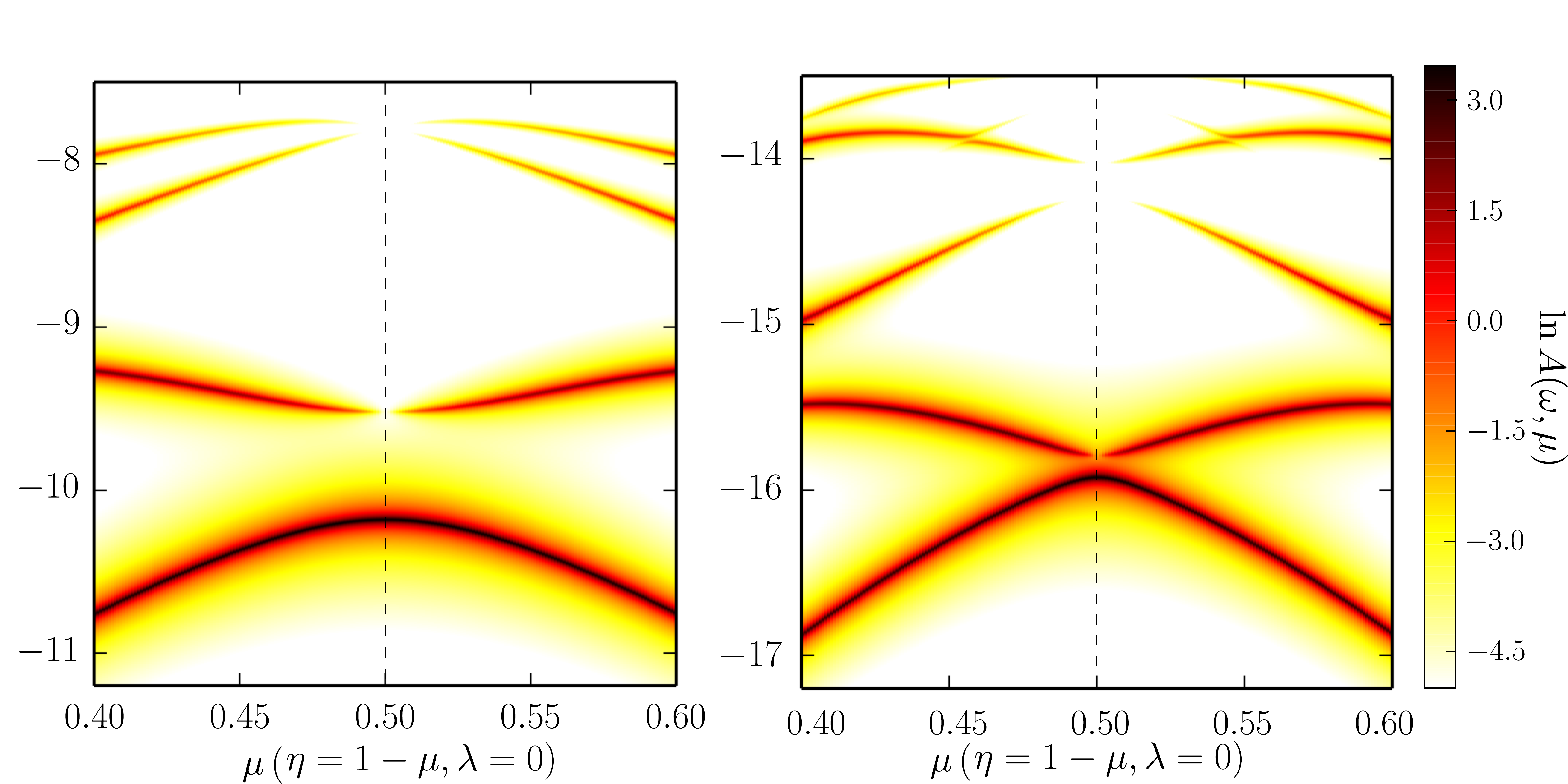}\caption{Spectral decomposition  $A$  of the ground state wavefunction at the transition points (dashed lines) for different system sizes. The left column corresponds to 48 spins and the right column correspond to 75 spins. 
The top row shows the behavior of $A(\omega,\lambda)$ at
the known second-order transition from the TC phase to the polarized
phase at $\lambda\approx0.18$. The bottom row shows $A(\omega,\mu)$
at the transition from the TC to the DSem phase where 
a transition occurs at $\mu=0.5$.}
\label{TransitionFidelity}
\end{figure}

Figures \ref{TransitionFidelity}(a) and \ref{TransitionFidelity}(b)
show $A(\omega,\lambda)$ for Hamiltonian (\ref{eq:mainH}) with $\mu=0$ along the line $\eta = 1 - \lambda$ for two different  system sizes. 
A second order  phase transition from the TC phase to a polarized phase takes place at  $\lambda_c\approx0.18$.\cite{Trebst2007} The location of this transition is known accurately since the TC can be mapped to an Ising model on a triangular lattice
(Appendix A) for which the transition has been studied in detail.\cite{Bloete2002}
The levels above the ground state become denser with increasing system size and 
appear to condense as expected in a second order transition.

Figures \ref{TransitionFidelity}(c) and \ref{TransitionFidelity}(d)
show $A(\omega,\mu)$ with $\lambda=0$ along the line $\eta = 1 - \mu$ for two different  system sizes. 
We observe a level crossing as the system size is increased. This again points to a first-order
transition at $\mu=\eta=0.5$.  Although we have frozen out electric charge defects, we note that
our conclusions about the phase transition of Hamiltonian (\ref{eq:mainH}) remain valid when allowing for 
electric defects as long as $\epsilon_{V}$ is chosen large enough. This is because the 
number and the positions of electric charges are good quantum numbers at all points in the phase diagram. 
The nature of the phase transition may alter if we add terms to the Hamiltonian which can create and hop electric 
defects (e.g., a transverse field). We defer the investigation of this issue to a possible future work. 
The nature of the transition between TC and DSem may indeed be sensitive to precisely what additional 
terms are present in the Hamiltonian. For instance, Ref.~\onlinecite{Barkeshli2013} notes that a second-order 
transition is possible in the presence of an additional $SU(2)$ symmetry in the Hamiltonian.

\subsection{Topological Entanglement Entropy(TEE)}

We now identify the region of the phase diagram that is topologically
ordered. To do this, we calculate the topological entanglement entropy\cite{Kitaev2006,Levin2006}
(TEE) which is widely used to identify topologically ordered phases.
The entanglement entropy of a subsystem is given by the von Neumann entropy
of its reduced density matrix. The TEE ($\gamma$) is a length-independent correction
to the area law for the entanglement entropy. In 2D, 
\begin{equation}
S=\alpha L-\gamma
\end{equation}
\noindent where L is the length of the boundary of the subsystem.

A topologically ordered phase has a non-zero TEE because of the long-range
entanglement in the system. For an abelian phase, the number of quasiparticles $n$
determines the TEE completely through  $\gamma=\log(\sqrt{n})$. The
TC and DSem phases are both abelian phases with four quasiparticles.
Hence, $\gamma=\log2$ for both phases. Thus, we expect that even
away from the fixed points, we should not be able to distinguish them
using TEE. As introduced in Ref. \onlinecite{Kitaev2006}, we calculate the TEE
numerically by adding and subtracting the entanglement entropy of
different regions such that the length dependent contributions cancel
out. 

\begin{figure}
\includegraphics[width=1\columnwidth]{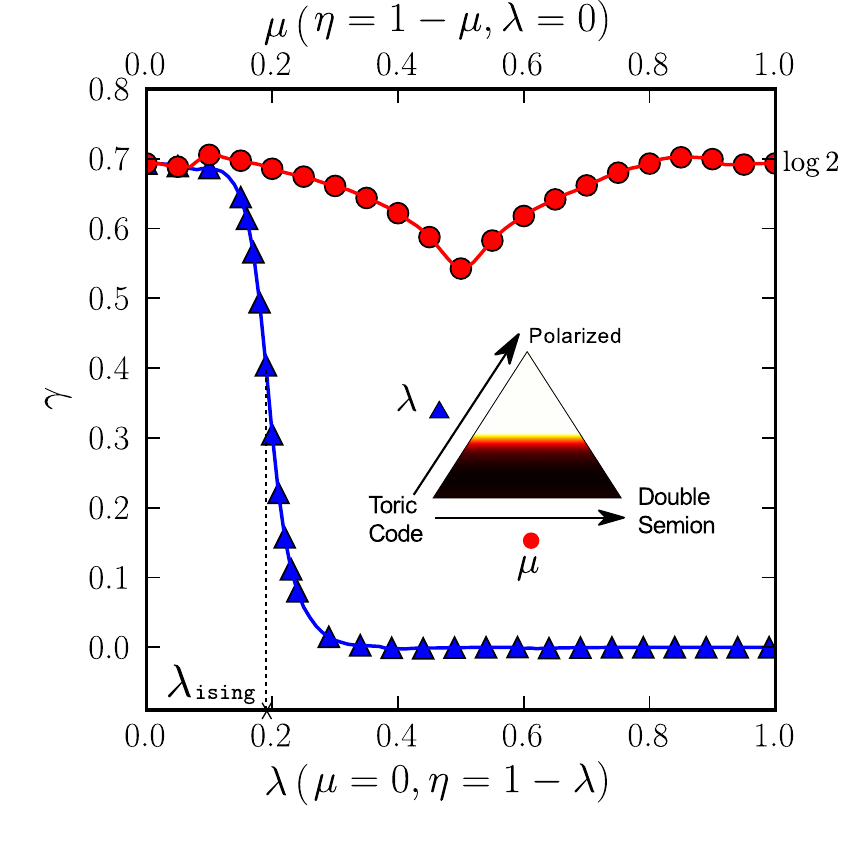}\caption{Variation of the topological entanglement entropy (TEE) as the system
is tuned from the TC fixed point to the DSem fixed point (along $\lambda=0$
- red circles, top x-axis) , and from the TC fixed point to the polarized
fixed point (along $\mu=0$ - blue triangles, bottom x-axis). The
inset shows the TEE at all points in the phase diagram with dark red
regions indicating higher TEE. This indicates that the bottom part
of the phase diagram is topologically ordered. The size of the system
considered is 48 spins.}
\label{TEE}
\end{figure}

Figure \ref{TEE} shows the TEE as the system is tuned along two different
directions in the phase diagram. The blue triangles show a transition
from the TC phase to a polarized phase\cite{Trebst2007} on increasing
the magnetic field. The red circles show a transition
from the TC phase to the DSem phase (with zero magnetic field). We
see a dip at the transition point $\mu=0.5$ due to finite-size effects,
but the TEE remains the same on both sides of the transition. The triangular
inset shows the TEE at all points in the phase diagram. The dark areas
indicate regions with high TEE and correspond to topologically
ordered phases with TEE close to $\log2$.

Hence, TEE  distinguishes the topological regions from the
trivial regions in the phase diagram. However, it cannot distinguish
between the different topologically ordered phases. To do that, we turn to extracting
the braiding statistics of the excitations in the topological region.

\subsection{Braiding Statistics}

\begin{figure}
\includegraphics[width=0.95\columnwidth]{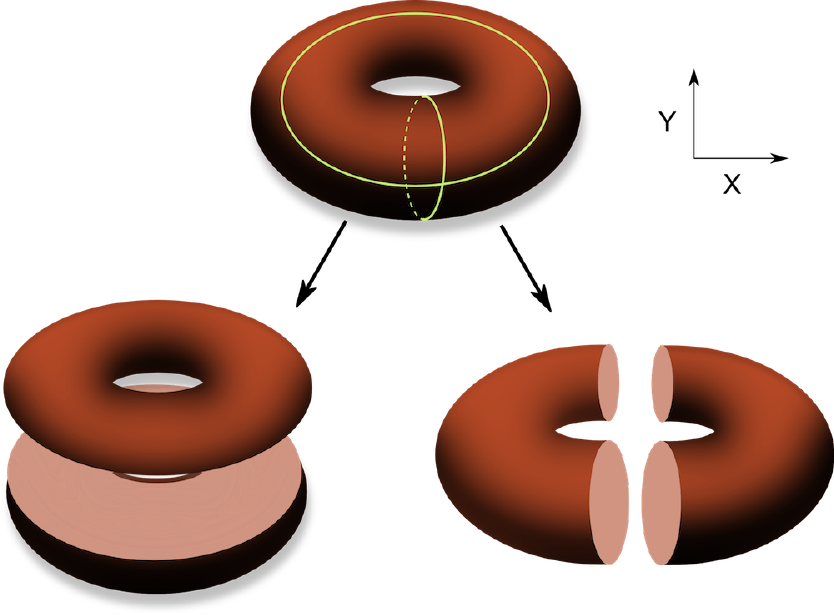}\caption{Non-trivial bipartitions of the torus. We consider the partition on
the right and trace over one half of the torus to obtain the reduced
density matrix. The yellow lines indicate the long loops around the torus
characterizing the possible winding sectors.}
\label{TorusPartition}
\end{figure}

We extract the braiding statistics of the excitations in the topologically
ordered region. In particular, we obtain the $U$ and $S$-matrices which
quantify the self statistics (exchange statistics) and the mutual statistics
respectively. An element of the $U$-matrix specifies the phase obtained by the many-body wavefunction 
when we exchange two identical particles.  An element of the $S$-matrix corresponds 
to the phase obtained when we move one of the 
particles in a closed path around another particle. 

The $U$ and $S$-matrices of the TC and DSem model can be calculated exactly.\cite{PhysRevB.71.045110,Zhang2012}
However, away from these fixed points, the Hamiltonian is no longer
exactly solvable and we have to obtain the matrices numerically.

We obtain the $U$ and $S$-matrices by constructing overlaps of so-called
Minimally Entangled States (MES) on non-trivial bipartitions of a torus\cite{Zhang2012,Cincio2013} (Fig.~\ref{TorusPartition}).
The MES are ground states which minimize the bipartite entanglement
entropy on a given bipartition of the system. When we consider non-trivial
bipartitions of a torus, it turns out that the MES are eigenstates
of Wilson loop operators defined parallel to the entanglement cuts.
Now, take for example the $S$-matrix. It is the matrix
that transforms the eigenstates of one Wilson loop operator to another.
Thus, it follows that the $S$-matrix corresponds to a certain unitary transformation in the MES basis. 

To obtain the MES, we need to first  calculate all the degenerate ground
states of the system. There are four ground states for both the TC and
the DSem phase on a torus. These can be characterized by their winding numbers
(modulo 2) around two independent directions of the torus, i.e., by having an even 
or odd number of loops winding around the torus. We  obtain four almost degenerate ground states
for the Hamiltonian of Eq.~(\ref{eq:mainH}) by diagonalizing it separately in each
winding number sector. We then obtain the MES ($|\Xi\rangle$) by minimizing
the entanglement entropy for a linear combination of these ground
states. Since we have four ground states, our parameter space consists
of the surface of a 3-sphere plus additional phase factors,
\begin{equation}
|\Xi^{x}\rangle=\xi_{1}|00\rangle+e^{i\phi_{1}}\xi_{2}|01\rangle+e^{i\phi_{2}}\xi_{3}|10\rangle+e^{i\phi_{4}}\xi_{4}|11\rangle,
\end{equation}

\noindent where $|\Xi^{x}\rangle$ corresponds to a MES on one of
the non-trivial bipartitions of the torus and $|\alpha\beta\rangle$ is the ground state in the $\alpha\beta$
winding sector.

Since the model is defined on the honeycomb lattice, the transformation of the MES
under a $2\pi/3$  rotation allows us to calculate the $US$ matrix
(Appendix B).\cite{Zhang2012} As shown in the appendix, we can also use the $US$
matrix to calculate the $U$ and $S$ matrices individually. As an
example, we indicate the $U$ and $S$-matrices obtained at two points on
the transition along the line $\lambda=0$ with $\eta=1-\mu$. At $\mu=0.25$, we obtain matrices close to the exact
ones for the TC model

\[
U_{jj}^{0.25}=\left(\begin{array}{c}
1.0\\
1.0\\
1.0\\
-1.0
\end{array}\right)+10^{-1}\left(\begin{array}{c}
1.5e^{-i0.1\pi}\\
0.7e^{-i0.1\pi}\\
1.0e^{i0.9\pi}\\
0.8e^{i0.1\pi}
\end{array}\right)
\]

\[
S^{0.25}=\frac{1}{2}\left(\begin{array}{cccc}
1.0 & 1.0 & 1.0 & 1.0\\
1.0 & 1.0 & -1.0 & -1.0\\
1.0 & -1.0 & 1.0 & -1.0\\
1.0 & -1.0 & -1.0 & 1.0
\end{array}\right)+
\]
\[
10^{-1}\left(\begin{array}{cccc}
0.3e^{i0.9\pi} & 0.3e^{i0.9\pi} & 0.3e^{i0.9\pi} & 0.3e^{i0.9\pi}\\
0.3e^{i0.9\pi} & 0.3e^{i0.7\pi} & 0.4e^{i0.6\pi} & 0.7e^{-i0.9\pi}\\
0.3e^{i0.9\pi} & 0.2e^{-i0.7\pi} & 0.5e^{-i0.1\pi} & 0.9e^{-i0.9\pi}\\
0.3e^{i0.9\pi} & 0.7e^{-i0.1\pi} & 0.1e^{i0.7\pi} & 0.4e^{i0.1\pi}
\end{array}\right)
\]

At $\mu=0.75$, we correspondingly obtain matrices close to the exact
ones for the DSem model.

\[
U_{jj}^{0.75}=\left(\begin{array}{c}
1.0\\
1.0\\
-1i\\
1i
\end{array}\right)+10^{-1}\left(\begin{array}{c}
1.3\\
0.7\\
0.8e^{i0.5\pi}\\
0.8e^{-i0.5\pi}
\end{array}\right)
\]
\[
S^{0.75}=\frac{1}{2}\left(\begin{array}{cccc}
1.0 & 1.0 & 1.0 & 1.0\\
1.0 & 1.0 & -1.0 & -1.0\\
1.0 & -1.0 & -1.0 & 1.0\\
1.0 & -1.0 & 1.0 & -1.0
\end{array}\right)+
\]
\[
10^{-1}\left(\begin{array}{cccc}
-0.3 & -0.3 & -0.3 & -0.3\\
-0.3 & -0.2 & 0.6e^{i0.8\pi} & 0.7e^{-i0.8\pi}\\
-0.3 & 0.7e^{-i0.3\pi} & 0.4e^{-i0.9\pi} & 0.7e^{i0.3\pi}\\
-0.3 & 0.7e^{i0.3\pi} & 0.7e^{-i0.3\pi} & 0.5e^{i0.9\pi}
\end{array}\right)
\]
\begin{figure}
\includegraphics[width=1\columnwidth]{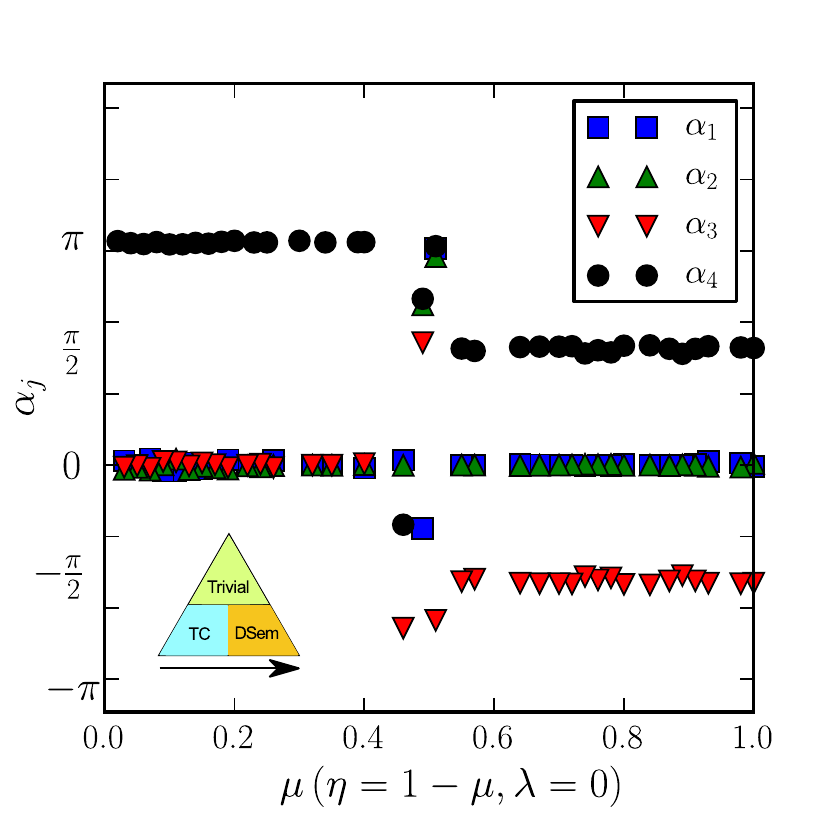}\caption{Variation of the $U$-matrix as the system is tuned from the TC phase
to the DSem phase. The angle $\alpha_{j}$ plotted is related to the
$U$-matrix by $U_{jj}=e^{i\alpha_{j}}$. We see a transition from the
fermionic statistics indicative of the TC phase to the semionic statistics
indicative of the DSem phase. The region near the transition point
is ambiguous due to an additional degeneracy which comes into play. The size
of the system considered is 48 spins.}
\label{UMatrix}
\end{figure}

From the plot of the elements of the $U$ matrix (Figure \ref{UMatrix}), we can identify 
the characteristic statistics of the quasiparticle excitations in the two phases. In the TC 
phase ($\mu<0.5$), three of the quasiparticles ($1$, $e$, $m$) have bosonic exchange statistics 
and one of them is a fermion ($\psi$). In the DSem phase ($\mu>0.5$), we find two semionic quasiparticles ($s^+$, $s^-$) 
and two of them behave like bosons ($1$, $m$). 

Near the transition point ($\mu=0.5$), the results obtained are ambiguous. This is possible due to two reasons.
First, minimizing the entanglement entropy is the same as maximizing the TEE. Hence, finding the correct MES depends 
on the four ground states having the same TEE. However, close to the transition point we observe that the TEE for the 4 states
becomes very different due to finite size effects and this causes a bias in the minimization. The second reason is that we have
 an extra degeneracy which arises at $\mu=0.5$ because the Hamiltonian conserves the loop number modulo 2. 
We then have two sectors characterized by the number of loops modulo 2 containing 4 topologically degenerate ground states each. 
We expect the two sectors to be degenerate in the thermodynamic limit and thus we end up with eight ground states at the transition point 
which violates our assumption about the parameter space of the minimization.  Despite this ambiguity due to finite size effects, we believe 
that there is a sharp transition as indicated previously from the energy spectrum.

Thus, by using the $U$ and $S$-matrices as non-local ``order parameters'', we uniquely characterize the topological region of the phase diagram by 
distinguishing the different topologically ordered phases in it.

\section{Summary and discussions\label{sec:Summary-and-discussions}}

In this paper, we considered a model Hamiltonian which exhibits two different $\mathbb{Z}_2$ topologically ordered phases and a  topologically trivial phase. 
In two limiting cases, the Hamiltonian reduces to the exactly solvable double semion (DSem) and toric code (TC) models.
We obtained an exact solution for the entire phase diagram on a 1D ladder and find a second-order transition half-way between the two limits, described
by an Ising$\times$Ising conformal field theory. 
For the 2D system on a honeycomb lattice, we resort to exact diagonalization in a basis of loops which allows us to go up to system sizes of 75 spins. 
We reproduce the known second order transition between the topological and trivial phases and find indications for a first-order transition between the two $\mathbb{Z}_2$ topological phases from the energy spectrum.
While the topological entanglement entropy distinguishes the topologically ordered and trivial phases, it cannot tell the difference between the TC and DSem phases.
We calculated the exchange statistics and mutual statistics of the quasiparticle excitations numerically and used them as ``order parameters'' to distinguish the two topologically ordered phases.  
We finally obtained the full phase diagram of the Hamiltonian of Eq.~(\ref{eq:mainH}) which is schematically shown in the inset in Fig.~\ref{UMatrix}.
We showed that the braiding statistics for different $\mathbb{Z}_2$ phases in a spin system can be obtained directly using exact diagonalization methods. 
The same method might be used in the case of the Heisenberg antiferromagnet on the kagome lattice.
Although the ground state is believed to contain $\mathbb{Z}_2$ topological order, \cite{Yan2011,Jiang2012,Depenbrock2012} a more precise characterization could be made by calculating the $U$ and $S$ matrices to determine whether it corresponds to a TC or DSem phase.

\begin{acknowledgments}
We thank Fiona Burnell, Andreas L{\"a}uchli, Adam Nahum, Kirill Shtengel,  Steve Simon, Ari Turner, and Mike Zaletel for helpful discussions. C. K. gratefully acknowledges the hospitality of the guest program of MPI-PKS Dresden.
\end{acknowledgments}

\appendix

\subsection{Appendix A: Mapping to an Ising model on a triangular lattice}

\begin{figure}
\includegraphics[width=0.95\columnwidth]{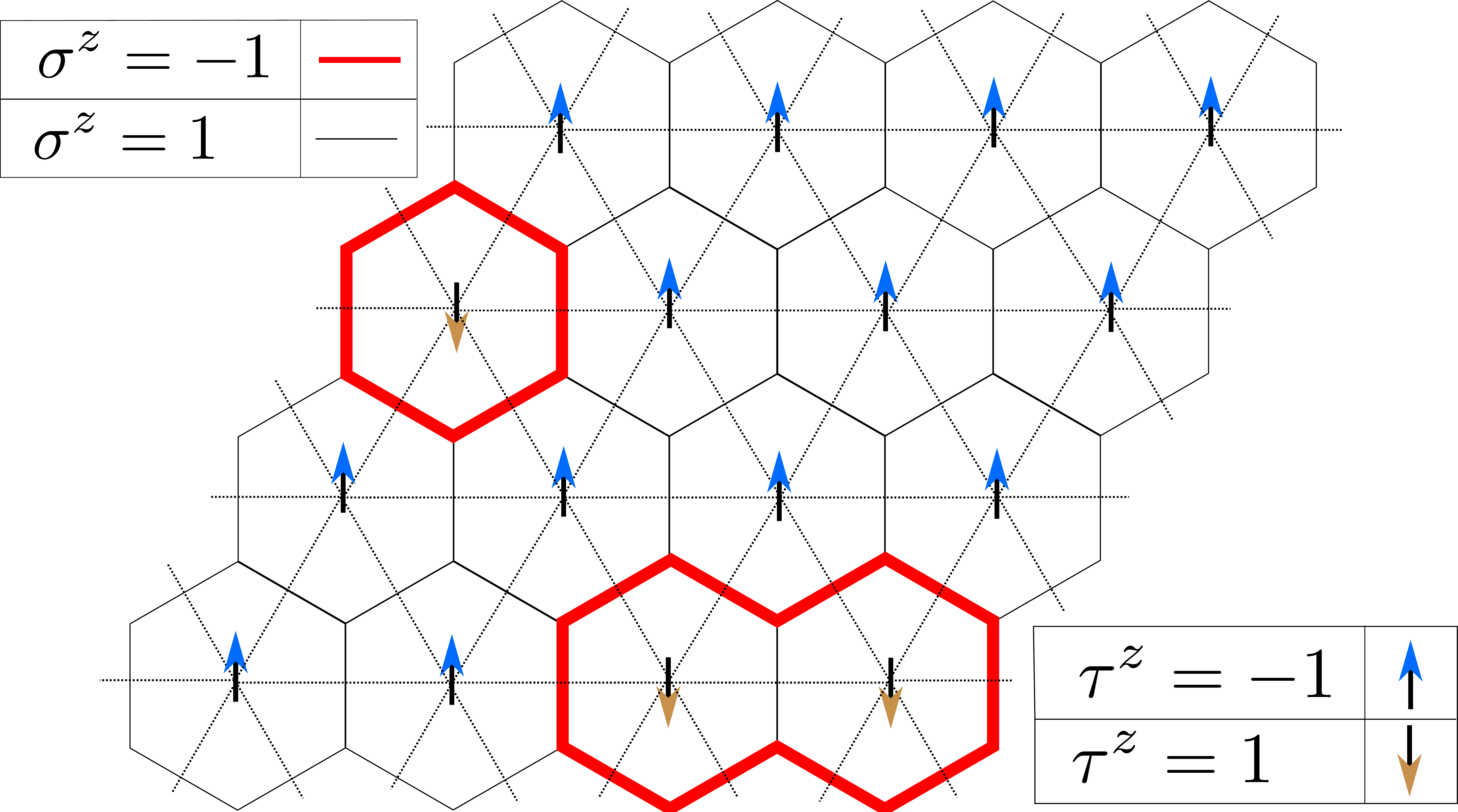}\caption{Duality mapping between the loops in the topological
models and domain walls in the dual $\tau_{p}^{z}$ Ising spins which
live on plaquettes. }
\label{TriangularMapping}
\end{figure}

Here we remind the reader of the simple duality mapping between the
topological TC/DSem models and their dual (non-topological) Ising-like models.
For simplicity, consider the TC or DSem model on a sphere. Let us
forbid vertex defects so that the Hilbert space consists only of all
possible loop configurations. There is a one-to-one mapping between
this Hilbert space and the space of all domain wall configurations
in Ising degrees of variables $\tau^{z}=\pm$1 living at the centers
of plaquettes (Figure \ref{TriangularMapping}). For example, in this language
we may rewrite the DSem Hamiltonian as 

\begin{equation}
H^{\mathrm{DSem}}=+\epsilon_{P}\!\!\sum_{p}\tau_{p}^{x}\prod_{\left\langle qr\right\rangle_{p} }i^{(1-\tau_{q}^{z}\tau_{r}^{z})/2}
\end{equation}

\noindent where $\left\langle qr\right\rangle_{p} $ is the set of plaquettes
$q,r$ which are nearest neighbors with one another, and nearest
neighbors with $p$. To ensure a one to one correspondence between
the Hilbert spaces of the two models, we also need to impose the global
symmetry $\prod_{p}\tau_{p}^{x}=1$. 

Notice that, as a whole, this dual Hamiltonian has Ising symmetry.
This is no coincidence. This Hamiltonian arose from a model that acted
on configurations of loops, which are now interpreted as domain walls.
This resulting Hamiltonian should only care about domain walls, rather
than the actual values of the Ising spins within the domains. Hence,
notice that the $\tau^{z}$ operators only ever arise in pairs, which
gives rise to a global Ising symmetry. There is related to the subtlety
that there is a one-to-two mapping between loop configurations and
the Ising spins themselves. This is because Ising spin configurations
related by flipping all the spins give rise to the same domain wall
configurations.  Another way of expressing this redundancy is to note 
that $\prod_{p} B_{p}\equiv1$ on the Hilbert space of the string-net models. 
In terms of the dual Ising spins, this relation reads $\prod_{p} \tau_{p}^{x} =1$; 
this does not hold as an identity, and needs to be imposed as a symmetry constraint.

\subsection{Appendix B: Obtaining braiding statistics from Minimally Entangled States (MES)}

The $U$ and $S$ matrices can be extracted from the MES by the method
outlined in Ref. \onlinecite{Zhang2012},\onlinecite{Cincio2013}.
We can construct the $US$ matrix by observing the action of a $2\pi/3$
rotation on the MES over a lattice which has a $2\pi/3$ symmetry,
\begin{equation}
(D^{\dagger}USD)_{ij}=\langle\Xi_{j}^{x}|R_{2\pi/3}|\Xi_{i}^{x}\rangle
\end{equation}

\noindent where $D$ is a diagonal matrix representing the unknown phases that
come with each $|\Xi^{x}\rangle$ 

And we can construct the $S$ matrix by observing the action of a
$\pi/2$ rotation on the MES on a lattice which has a $\pi/2$ symmetry,
\begin{equation}
(S)_{ij}=\langle\Xi_{j}^{x}|R_{\pi/2}|\Xi_{i}^{x}\rangle
\end{equation}

However, we cannot extract the $S$ matrix this way since our honeycomb
lattice does not have a $\pi/2$ symmetry. But, we can extract the
$U$ and $S$ matrix from the $US$ matrix by using some of the constraints
on them. Since we know that the $U$ and $D$ matrices are diagonal,
we have 24 unknown parameters (16 from the $S$ matrix, 4 from the
$U$ matrix and 4 phases). However, we only have 16 constraints from
the action of the $2\pi/3$ rotation on the MES. 

We can get around this by assuming that we know which of the degenerate 
ground states the identity particle (which always exists) belongs to. We will verify this assumption
later. Now we know that the identity particle has trivial self statistics
($U_{1}=1$) and trivial mutual statistics ($S_{1i}=S_{i1}=\frac{1}{Q}$
where $Q$ is the quantum dimension). Since we have an
abelian phase with 4 quasi-particles,  $Q=2$ here. We can also obtain it from the
topological entanglement entropy ($\gamma$) by using $Q=e^{\gamma}$.
Thus, we reduce the number of unknown parameters to 16. We can now
solve for all the parameters.

In fact, we actually have more constraints since we also know that
the $S$-matrix is symmetric. We can test our assumption about the identity
particle by checking that the $S$-matrix we obtain is symmetric and
that the row and column of the $S$-matrix and the element of the $U$-matrix corresponding 
to the identity particle is approximately equal to one. We do this along all
the points in our phase diagram and find an approximately symmetric
$S$-matrix everywhere, thus validating our original assumption. 

%

\end{document}